\documentclass[12pt]{article}
\usepackage{amsfonts,amsmath,amssymb}
\usepackage{amsthm,amscd}
\usepackage{subeqnar}
\makeatletter
\renewcommand{\theequation}{\thesection.\arabic{equation}}
\@addtoreset{equation}{section}
\makeatother
\everymath{\displaystyle}
\theoremstyle{definition}
\newtheorem{thm}{Theorem}

\newtheorem{pr}[thm]{Proposition}
\newtheorem*{defn}{Definition}
\newtheorem*{rem}{Remark}

\newtheorem*{pproof}{Proof}

\newenvironment{namelist}[1]{%
  \begin{list}{}
    {
    \settowidth{\labelwidth}{#1}
    \setlength{\leftmargin}{1.1\labelwidth}}
}{%
\end{list}}

\newcommand{\bibun}[2]{\frac{d {#1}}{d {#2}}}
\newcommand{\bib}[2]{\frac{\partial {#1}}{\partial {#2}}}

\newcommand{\bvec}[1]{\mbox{\boldmath $#1$}}

\begin{document}
\title{Spin(7) holonomy manifold and Superconnection}
\author{
\thanks{E-mail address : yasui@sci.osaka-cu.ac.jp} Yukinori YASUI  and
\thanks{E-mail address : ootsuka@sci.osaka-cu.ac.jp} Takayoshi OOTSUKA \\
{\small Department of Physics, Osaka City University, 
Sumiyoshiku, Osaka, Japan } }
\date{}
\maketitle


\begin{abstract}

We discuss the higher dimensional generalization
of gravitational instantons by using volume-preserving
vector fields.
We give special attention to the case of 8-dimensions 
and present a new construction of the Ricci flat metric
with holonomy in Spin(7).
An example of the metric is explicitly given.
Further it is  shown that our formulation has a natural interpretation
in the Chern-Simons theory written by the language of
superconnections.
\quad \\
\quad \\
\end{abstract}


\section{Introduction}

Higher dimensional gravitational instantons play a prominent role in 
our understanding of soliton solutions in supergravities and 
superstrings.
Indeed, in~\cite{D-P-S, G-G-P-T, B-F-K, B-P}, 
these instantons have been interpreted as
D-branes in ten or eleven dimensional space-time.
Geometrically, gravitational instantons are intimately linked to
Riemannian manifolds with special holonomy groups.
In this paper we will mainly study Spin(7) holonomy manifolds
that correspond to 8-dimensional gravitational instantons
although a brief discussion for ${\rm G}_2$ holonomy manifolds
(7-dimensional gravitational instantons) will be given.
The holonomy groups Spin(7) and ${\rm G}_2$ 
constitute the two exceptional members of 
Berger's list of holonomy groups~\cite{Besse}.
In common with the other groups SU($n$) and Sp($n$) in the list
these exceptional holonomy manifolds have the property that their
metrics are automatically Ricci-flat.

It is well known that in 4-dimensions gravitational instantons
are the solutions to the self-duality equation;
the corresponding Riemannian manifolds are hyperk\"ahler manifolds
with holonomy group Sp(1).
The 4-dimensional self-duality condition on the curvature 2-form 
$R_{\alpha \beta}$ can be generalized to 
\begin{eqnarray}
  R_{\alpha \beta} = \frac12 \Omega_{\alpha \beta \gamma \delta}
  R_{\gamma \delta}
\end{eqnarray}
where $\Omega_{\alpha \beta \gamma \delta}$ is a duality operator,
identified with the components of a certain
4-form (Cayley 4-form) on an 8-dimensional manifold~\cite{Acha, B-F-K}.
The manifolds with this "self-duality" condition have holonomy in
Spin(7) and can be thought of as 8-dimensional gravitational
instantons.

In 4-dimensions, there is another construction of gravitational
instantons (hyperk\"ahler manifolds) based on Ashtekar gravity.
In this approach the instantons are given by the solutions to the 
differential equations for volume-preserving vector fields.
The following proposition summarizes the results of
~\cite{A-J-S, M-N, Don, Hashi}
relevant to this construction. 

\pr
Let $(M, \omega)$ be a 4-dimensional manifold with volume form
$\omega$ and $V_\mu \, (\mu =0,1,2,3)$ be linearly independent
four vector fields satisfying the conditions:
\begin{eqnarray}\label{eq:MN}
 {\bf (a)} \quad  L_{V_\mu} \omega = 0, \qquad \qquad
 {\bf (b)} \quad {\bar{\eta}^a}_{\mu \nu} [V_\mu, V_\nu ] = 0,
\end{eqnarray}
where ${\bar{\eta}^a}_{\mu \nu}\, (a=1,2,3)$ are the 't Hooft matrices
defined as
\begin{eqnarray}
  {\bar{\eta}^a}_{ij} = \epsilon_{aij}, \quad 
  {\bar{\eta}^a}_{0i} = \delta_{ai} \quad (a,i,j=1,2,3), \quad
  {\bar{\eta}^a}_{\mu \nu} = - {\bar{\eta}^a}_{\nu \mu}.
\end{eqnarray}
Then these vector fields induce a hyperk\"ahler metric on $M$:
\begin{eqnarray}
  g= \phi~W^\mu \otimes W^\mu, \quad
  \phi = \omega(V_0,V_1,V_2,V_3),
\end{eqnarray}
where $W^\mu$ is the dual basis of $V_\mu$ and the 2-forms
\begin{eqnarray}
  \Sigma^a= \frac12 \phi~{\bar{\eta}^a}_{\mu \nu} W^\mu
  \wedge W^\nu \, (a=1,2,3)
\end{eqnarray}
give the hyperk\"ahler forms.
\vspace{12pt}

This proposition yields that the vector fields $V_\mu$ may be
identified with the components of a spacetime-independent
Yang-Mills connection on $\Bbb{R}^4$. 
Indeed, the condition (a) asserts that the gauge group 
is the diffeomorphism
group on $M$ preserving the volume form $\omega$, and 
the condition (b) is 
equivalent to the reduced anti-self-dual Yang-Mills equation.

Now we can ask the question whether it is possible to generalize
the proposition 1 to the case of Spin(7) holonomy manifolds.
The purpose of this paper is to give an answer to this question.
In Section 2 we provide a new construction of Spin(7)
holonomy manifolds including the brief review of the
geometry for these manifolds.
Our formulation actually gives an 8-dimensional generalization
of the Proposition 1, and the result is summarized in the
Proposition 2.
As an example we also present a metric with holonomy Spin(7)
on the 8-dimensional space $\Bbb{R}^2 \times S^3 \times S^3$.
This is the same as the metric constructed by using different 
methods~\cite{B-S, G-P-P}.
In Section 3 we give an interpretation of the Proposition 2
from the viewpoint of 8-dimensional Yang-Mills theory
(generalized Chern-Simons theory).
We then formulate the theory using the superconnection
introduced by Quillen to describe the characteristic classes
of fiber bundles.
The solutions to the equations of motion are directly related
to the metrics of Spin(7) holonomy manifolds by 0-dimensional
reduction.
Section 4 is devoted to discussion.
We argue the application of our formulation to ${\rm G}_2$
holonomy manifolds and also make a remark on higher dimensional
Ashtekar gravity.

\section{Spin(7) holonomy manifold}

We begin by reviewing some of the properties of Spin(7) holonomy
manifold $M$ that we shall need in this
paper~\cite{G-P-P, Harvey}.
Such a manifold is 8-dimensional and admits a covariant constant
Majorana-Weyl spinor $\zeta$ invariant under the action of Spin(7).
It is interesting to note that the existence of $\zeta$ automatically
implies that the manifold $M$ must be Ricci-flat.
According to the standard isomorphism between the space of forms
and tensor product of the Clifford module, 
one can immediately construct a 4-form $\Omega$ on $M$,
with components given by
\begin{eqnarray}
  \Omega_{\alpha \beta \gamma \delta} = 
  \zeta^T \Gamma_{\alpha \beta \gamma \delta} \zeta,
\end{eqnarray}
where $\Gamma_{\alpha \beta \gamma \delta}$ denotes the
anti-symmetrized 4-fold product of the $\gamma$-matrices
$\Gamma_{\alpha}$ in 8-dimensions.
This 4-form enjoys a couple of remarkable properties;
(a) closedness, (b) self-duality with respect to usual Hodge 
star operator, (c) Spin(7)-invariance.
Conversely, a manifold has holonomy in Spin(7)
if there exists a 4-form $\Omega$ satisfying the conditions (a), (b)
and (c).
Such a 4-form $\Omega$ is known as the Cayley 4-form.
\subsection{Cayley 4-form}
The purpose of this section is to study Spin(7) holonomy manifolds
by using volume-preserving vector fields.
We will see that these vector fields induce the Cayley 4-form on 
an 8-dimensional space.

Before stating our proposition, we give a brief exposition of the
Cayley 4-form
\begin{eqnarray}\label{flat-Cayley}
  \Omega = \frac{1}{4!}\Omega_{\alpha \beta \gamma \delta}
  dx^\alpha \wedge dx^\beta \wedge dx^\gamma \wedge dx^\delta
\end{eqnarray}
on the Euclidean space $\Bbb{R}^8=\{ (x^1, x^2, \cdots , x^8)\}$.
A more detailed treatment may be found in~\cite{D-F-T, Harvey}.
The components $\Omega_{\alpha \beta \gamma \delta}$ are related to
the structure constants of octonionic algebra ${\cal O}$.
A basis for ${\cal O}$ is provided by the eight elements 
$1, \, e_a$ ($a=1,2,\cdots , 7$) which satisfy the relation
\begin{eqnarray}
  e_a e_b = \varphi_{abc} e_c - \delta_{ab}.
\end{eqnarray}
The structure constant $\varphi_{abc}$ is totally anti-symmetric with
\begin{eqnarray}\label{varphi}
  \varphi_{abc}=1 \qquad (abc)=(123),(516),(624),(435),(471),(673),(572).
\end{eqnarray}
Then $\Omega_{\alpha \beta \gamma \delta}$ is given by 
\begin{eqnarray}\label{Omega}
  \Omega_{abc8}=\varphi_{abc}, \quad \Omega_{abcd}=
  \frac{1}{3!} \epsilon_{abcdefg}\varphi_{efg}.
\end{eqnarray}
The 4-form $\Omega$ is obviously self-dual and, in addition, 
it is invariant under the action of Spin(7).
The last property can be confirmed as follows.
For the space of 2-forms $\wedge^2(\Bbb{R}^8) \cong \frak{so}(8)$,
the decomposition in irreducible Spin(7)-modules is given by
\begin{eqnarray}
  \wedge^2(\Bbb{R}^8)= \wedge^2_+ \oplus \wedge^2_-,
\end{eqnarray}
and the dimensions turn out to be $\mbox{dim}\wedge^2_+=7$
and $\mbox{dim}\wedge^2_-=21$.
$\wedge^2_-$ is isomorphic to $\frak{spin}(7)$ and the projection is
explicitly written as
\begin{eqnarray}
  M_{\alpha \beta}^- = \frac34 
  (M_{\alpha \beta} + \frac16 \Omega_{\alpha \beta \gamma \delta}
  M_{\gamma \delta}) \in \wedge^2_-.
\end{eqnarray}
If we identify $M_{\alpha \beta}$ with the 
standard generators of SO(8),
it is easy to show that the Spin(7) generators 
$M_{\alpha \beta}^-$ leave the 4-form $\Omega$
invariant.

Now we proceed to our main result,
which is a generalization of the Proposition 1 to
Spin(7) holonomy manifolds.

\pr
Let $(M, \omega)$ be an 8-dimensional manifold with volume form
$\omega$.
Let $V_\alpha \, (\alpha=1,2,\cdots,8)$ be linearly independent eight 
vector fields on $M$ that satisfy the following two conditions:
\begin{description}
\item [(a)] volume-preserving condition
  \begin{eqnarray}\label{eq:8vol-pre}
    L_{V_\alpha} \omega = 0,
  \end{eqnarray}
\item [(b)] 2-vector condition 
\begin{eqnarray}\label{eq:8MN}
  \Omega_{\alpha \beta \gamma \delta} [V_\alpha \wedge V_\beta ,
  V_\gamma \wedge V_\delta ]_{\rm SN} = 0,
\end{eqnarray}
\end{description}
where $\Omega_{\alpha \beta \gamma \delta}$ are the coefficients
defined by (\ref{Omega}) and $[~ , ~]_{\rm SN}$ is provided by the
Schouten-Nijenhuis bracket (see Appendix).
Then we have a Ricci-flat metric 
\begin{eqnarray}
  g = \phi~W^\alpha \otimes W^\alpha, \quad 
  \phi = \sqrt{\omega(V_1,V_2,\cdots,V_8)},
\end{eqnarray}
where $W^\alpha$ is the dual basis of $V_\alpha$ 
and 
\begin{eqnarray}
  \Omega = \frac{1}{4!} \phi^2~\Omega_{\alpha \beta \gamma \delta}
  W^\alpha \wedge W^\beta \wedge W^\gamma \wedge W^\delta
\end{eqnarray}
gives the Cayley 4-form on $M$.

\pproof 
If we introduce the 1-form 
$E^\alpha = \sqrt{\phi} W^\alpha$, then the 4-form $\Omega$
takes the form,
\begin{eqnarray}\label{eq:Omega}
  \Omega = \frac{1}{4!} \Omega_{\alpha \beta \gamma \delta}
  E^\alpha \wedge E^\beta \wedge E^\gamma \wedge E^\delta
\end{eqnarray}
and the metric is orthonormal 
in this frame, $g=E^\alpha \otimes E^\alpha$.
Thus a completely analogous result to the $\Bbb{R}^8$
case holds if we replace the 1-form $dx^\alpha$ with
$E^\alpha$, i.e. $\Omega$ is a Spin(7) invariant self-dual
4-form.
We shall now prove $\Omega$ is closed.
For this we rewrite the equation (\ref{eq:Omega})
in the form
\begin{eqnarray}
  \Omega = \frac{1}{4!} \Omega_{\alpha \beta \gamma \delta}~
  \iota_{V_\alpha} \iota_{V_\beta} \iota_{V_\gamma}
  \iota_{V_\delta} \omega,
\end{eqnarray}
where $\iota_{V_\alpha}$ denotes the inner derivation
with respect to $V_\alpha$.
By using the following formulas successively,
\begin{eqnarray}
  L_{V_\alpha} \iota_{V_\beta} - \iota_{V_\beta} L_{V_\alpha}
  = \iota_{[V_\alpha , V_\beta ]}, \quad 
  L_{V_\alpha}=d \iota_{V_\alpha} + \iota_{V_\alpha} d
\end{eqnarray}
we find 
\begin{eqnarray}
  d \Omega = \frac{1}{4!} \Omega_{\alpha \beta \gamma \delta}
  (6 \iota_{[V_\alpha , V_\beta ]}\iota_{V_\gamma} \iota_{V_\delta}
  \omega - 4 \iota_{V_\alpha}
  \iota_{V_\beta} \iota_{V_\gamma} L_{V_\delta} \omega
  + \iota_{V_\alpha}
  \iota_{V_\beta} \iota_{V_\gamma} \iota_{V_\delta} d \omega),
\end{eqnarray}
and furthermore the volume-preserving condition 
$L_{V_\alpha} \omega = 0$ with $d \omega = 0$
simplifies the equation,
\begin{eqnarray}
  d \Omega = \frac{1}{4!} \Omega_{\alpha \beta \gamma \delta}~
  \iota_{[V_\alpha , V_\beta ]}\iota_{V_\gamma} \iota_{V_\delta}
  \omega.
\end{eqnarray}
Finally, noting the identity
\begin{eqnarray}
  [V_\alpha \wedge V_\beta , V_\gamma \wedge V_\delta ]_{\rm SN}
  = [V_\alpha , V_\gamma ] \wedge V_\beta \wedge V_\delta
  - [V_\alpha , V_\delta ] \wedge V_\beta \wedge V_\gamma 
  \nonumber \\
  - [V_\beta , V_\gamma ] \wedge V_\alpha \wedge V_\delta 
  + [V_\beta , V_\delta ] \wedge V_\alpha \wedge V_\gamma, 
  \label{SN}
\end{eqnarray}
and the 2-vector condition (\ref{eq:8MN}),
we see that $\Omega$ is closed.
Thus $\Omega$ is the Cayley 4-form on $M$,
and $g$ a Ricci-flat metric with the holonomy in Spin(7).

\rem
If $V_\alpha$ is a solution to (\ref{eq:8vol-pre}) and 
(\ref{eq:8MN}),
then an infinitesimal deformation by the volume-preserving
vector field $X$,
\begin{eqnarray}\label{eq:GT}
  \tilde{V}_\alpha = V_\alpha + \varepsilon [X, V_\alpha]
\end{eqnarray}
gives a new solution.
Indeed, we have $L_{\tilde{V}_\alpha}\omega=0$ and
\begin{eqnarray}
  \lefteqn{
  \Omega_{\alpha \beta \gamma \delta} 
  [\tilde{V}_\alpha \wedge \tilde{V}_\beta , 
  \tilde{V}_\gamma \wedge \tilde{V}_\delta ]_{\rm SN}
  = \Omega_{\alpha \beta \gamma \delta}
  [V_\alpha \wedge V_\beta , V_\gamma \wedge V_\delta ]_{\rm SN}
  \nonumber }\\ && +
  \varepsilon \Omega_{\alpha \beta \gamma \delta}( 
  [[X, V_\alpha \wedge V_\beta]_{\rm SN}, 
  V_\gamma \wedge V_\delta]_{\rm SN} + 
  [V_\alpha \wedge V_\beta , 
  [X, V_\gamma \wedge V_\delta]_{\rm SN} ]_{\rm SN}),
\end{eqnarray}
which vanishes by (\ref{eq:8MN}) and the super Jacobi 
identity (\ref{eq:sJacobi}) up to the order $\varepsilon$.
This symmetry will be properly realized in Section 3 as 
the gauge transformations of superconnection.
It should be also noticed that (\ref{eq:GT})
induces the metric deformation $g \to g + L_X g$,
and hence this deformation can be absorbed in the 
coordinate transformation. 
\subsection{Example}
We illustrate our formulation by concentrating on
the 8-dimensional space $M=\Bbb{R}^2\times S^3 \times S^3$.
Let $(X,Y)$ be the natural coordinates on $\Bbb{R}^2$ and 
choose a volume form 
\begin{eqnarray}
  \omega = dX \wedge dY \wedge \Sigma^1 \wedge \Sigma^2 
  \wedge \Sigma^3 \wedge \sigma^1 \wedge \sigma^2 \wedge
  \sigma^3,
\end{eqnarray}
where $\Sigma^i$ and $\sigma^i$ ($i=1,2,3$) are SU(2) 
left-invariant 1-forms on the three-spheres satisfying 
the relations,
\begin{eqnarray}
  d\Sigma^i = - \frac12 \epsilon_{ijk} \Sigma^j 
  \wedge \Sigma^k , \quad 
  d\sigma^i = - \frac12 \epsilon_{ijk} \sigma^j
  \wedge \sigma^k.
\end{eqnarray}
We take the following ansatz for vector fields 
$V_\alpha \quad (\alpha=1,2,\cdots ,8)$ on $M$:
\begin{subeqnarray}\label{8V-Ansatz}
  V_i &=& a(X) (\alpha_{11}(Y) \Theta_i + \alpha_{12}(Y) \theta_i)
  \quad (i=1,2,3), \\
  V_{\hat{i}}&=&b(X) (\alpha_{21}(Y) \Theta_i + \alpha_{22}(Y) \theta_i)
  \quad (\hat{i}=4,5,6), \\
  V_7 &=& \beta(Y) \bib{}{X}, \\
  V_8 &=& c(X) \bib{}{Y},
\end{subeqnarray}
where $\Theta_i$ and $\theta_i$ are the dual basis of $\Sigma^i$
and $\sigma^i$, respectively.
Then these vector fields obviously preserve the volume form $\omega$,
i.e. $L_{V_\alpha} \omega = 0$.
It is straightforward to compute the Schouten-Nijenhuis bracket
$[V_\alpha \wedge V_\beta , V_\gamma \wedge V_\delta ]_{\rm SN}$
by the formula (\ref{SN}).
Substituting (\ref{8V-Ansatz}) into (\ref{eq:8MN}), 
we find 
\begin{eqnarray}
  \alpha_{11} =2, \quad  \alpha_{12} = 1-{\rm tanh}(Y),
  \quad \alpha_{21} = 0, \quad \alpha_{22}= \beta = {\rm sech}(Y),
\end{eqnarray}
and 
\begin{eqnarray}\label{bibun}
  \bibun{a}{X}=\frac12 \left( \frac{a^3}{b}-ab\right), \quad
  \bibun{b}{X}=-2a^2, \quad a=c.
\end{eqnarray}
Applying the Proposition 2,
we obtain a metric of Spin(7) holonomy
\begin{eqnarray}\label{metric}
  ds^2 = \sqrt{a^4b^3} dX^2 + \sqrt{b^3} {\rm sech}^2(Y)
  \left(dY^2 +\frac14 (\Sigma^i)^2\right) +\sqrt{\frac{a^4}{b}}
  \left(\sigma^i + \frac{1-{\rm tanh}(Y)}{2}\Sigma^i\right)^2
\end{eqnarray}
This metric was originally obtained by Bryant-Salamon~\cite{B-S},
Gibbons-Page-Pope~\cite{G-P-P} and further discussed in~\cite{B-F-K}.
We finish this section with some remarks:
\begin{description}
\item[(1) Coordinate transformation]
\quad \\
If we introduce the new variables
\begin{eqnarray}
  a = -\frac{3}{10}\left(\frac{9}{20}\right)^{\frac16}r^{\frac43}
  f^{-1}(r), \quad 
  b=\left(\frac{9}{20}\right)^\frac23 r^\frac43,
\end{eqnarray}
with the coordinate transformation
\begin{eqnarray}
  \bibun{X}{r}= -\frac{10}{3}\left(\frac{20}{9}\right)^\frac23
  r^{-\frac73} f^2(r),
\end{eqnarray}
then the equation (\ref{bibun}) reduces to
\begin{eqnarray}
  r\bibun{f}{r}= \frac53 (1-f^2) f.
\end{eqnarray}
The solution is explicitly given by
\begin{eqnarray}
  f(r)=\frac{1}{\sqrt{1-\left(
        \displaystyle{\frac{m}{r}}\right)^\frac{10}{3}}}~,
\end{eqnarray}
which reproduces the metric given in~\cite{G-P-P}.
\item[(2) Comparison with the Eguchi-Hanson metric]
\quad \\
Using the Proposition 1, 
we can construct the Eguchi-Hanson metric
from the vector fields $V_\mu \, (\mu=0,1,2,3)$ on 
$M=\Bbb{R} \times S^3$:
\begin{eqnarray}\label{4V-Ansatz}
  V_0 = \bib{}{X},\quad V_1=a(X) \Theta_1,\quad  V_2=a(X) \Theta_2,
  \quad V_3=b(X) \Theta_3
\end{eqnarray}
with
\begin{eqnarray}
  \bibun{a}{X}= ab, \quad \bibun{b}{X}= a^2.
\end{eqnarray}
Indeed, these vector fields preserve the volume form
$
\omega = dX \wedge \Sigma^1 \wedge \Sigma^2 \wedge \Sigma^3
$
and satisfy the condition 
${\bar{\eta}^a}_{\mu \nu} [V_\mu, V_\nu ] = 0$.
Thus we may regard the ansatz (\ref{8V-Ansatz}) as an 
8-dimensional analogue of (\ref{4V-Ansatz}).
Spin(7) manifolds have been extensively studied by Joyce~\cite{Joy}.
However, nothing is actually known about the explicit
metric except for (\ref{metric}).
On the other hand, we have many other examples
in 4-dimensions, Taub-NUT metric, Gibbons-Hawking metric, 
Atiyah-Hitchin metric and so on~\cite{E-G-H, A-H}.
It is interesting to investigate the Spin(7) holonomy metric
starting from more general ansatz.
We leave this issue to future research.
\end{description}

\section{Superconnection and Chern-Simons theory}
As mentioned in Section 1, there is a close connection between 
4-dimensional hyperk\"ahler manifolds and self-dual Yang-Mills
theories.
It is natural to ask whether any relation exists between Spin(7)
holonomy manifolds and 8-dimensional Yang-Mills theories.
The answer to this problem is positive;
we rewrite the equations (\ref{eq:8vol-pre}), 
(\ref{eq:8MN}) using
the language of superconnections,
and arrive at the equations of motion whose solution gives
a stational point of the (generalized) Chern-Simons action.

Let us recall Quillen's concept of a 
superconnection~\cite{Q, B-G-V}.
If $M$ is a manifold and $V=V^{+}\oplus V^{-}$ is a $\Bbb{Z}_2$-graded
vector space, let ${\cal A}(M, {\rm End}(V))$ be the space of 
${\rm End}(V)$-valued differential forms on $M$.
On the superalgebra ${\rm End}(V)$
\footnote{The algebra of endmorphisms ${\rm End}(V)$ is a
  superalgebra, when graded in the usual way:
${\rm End}^{+}(V)={\rm Hom}(V^+, V^+)\oplus {\rm Hom}(V^-,V^-),
\, {\rm End}^-(V)={\rm Hom}(V^+, V^-)\oplus {\rm Hom}(V^-,V^+)$},
there is a canonical commutator satisfying the axioms of the
Lie superalgebra (see Appendix):
\begin{eqnarray}
  [a, b] = a b - (-1)^{|a||b|} b a, \quad a, b \in {\rm End}(V).
\end{eqnarray}
The space ${\cal A}(M, {\rm End}(V))$ has a $\Bbb{Z}$-grading 
${\cal A}^n(M, {\rm End}(V))$ given by the degree of differential 
forms.
In addition, we have the total $\Bbb{Z}_2$-grading,
which we will denote by
\begin{eqnarray}
 {\cal A}(M, {\rm End}(V)) = {\cal A}^+(M, {\rm End}(V)) \oplus
  {\cal A}^-(M, {\rm End}(V)),
\end{eqnarray} 
where
\begin{eqnarray}
{\cal A}^{\pm}(M, {\rm End}(V))=
  \sum_k {\cal A}^{2k}(M, {\rm End}^{\pm}(V)) \oplus 
  \sum_k {\cal A}^{2k+1}(M, {\rm End}^{\mp}(V)).
\end{eqnarray}

\defn
A superconnection (super covariant derivative) on 
${\cal A}(M, {\rm End}(V))$ is an odd-parity first-order differential
operator
\begin{eqnarray}
  {\rm D}_{\bvec{A}} : {\cal A}^{\pm}(M, {\rm End}(V)) \longrightarrow
  {\cal A}^{\mp}(M, {\rm End}(V))
\end{eqnarray}
which satisfies Leibniz's rule in $\Bbb{Z}_2$-graded sense:
if $\alpha$ denotes an ordinary differential form and 
$\theta \in {\cal A}(M, {\rm End}(V))$ then
\begin{eqnarray}
  {\rm D}_{\bvec{A}} (\alpha \wedge \theta)= d \alpha \wedge
  \theta + (-1)^{|\alpha|} \alpha \wedge {\rm D}_{\bvec{A}}
  \theta.
\end{eqnarray}
In order to better understand the content of the superconnection,
we may write 
$$
  {\rm D}_{\bvec{A}} = d + \bvec{A}
$$
and 
$$
\bvec{A} = \bvec{A}^{[0]} + \bvec{A}^{[1]} + \bvec{A}^{[2]} + \cdots
$$
Here, $\bvec{A} \in {\cal A}^-(M, {\rm End}(V))$ 
and $\bvec{A}^{[k]}$ lies
in ${\cal A}^k(M, {\rm End}^-(V))$ if $k$ is even, 
and in ${\cal A}^k(M, {\rm End}^+(V))$
if $k$ is odd.

The supercurvature  defined by
\begin{eqnarray}
  \bvec{F} = d \bvec{A} + \bvec{A} \wedge \bvec{A} 
  \, \in {\cal A}(M, {\rm End}(V))
\end{eqnarray}
has total degree even, and satisfies the Bianchi identity
\begin{eqnarray}
  d \bvec{F} + \bvec{A} \wedge \bvec{F} - \bvec{F} \wedge
  \bvec{A} = 0.
\end{eqnarray}
For a Spin(7) manifold $M$ with Cayley 4-form $\Omega$,
we consider the Chern-Simons action,
\begin{eqnarray}
  S = \int_M \Omega \wedge {\rm str}(\bvec{A} \wedge d \bvec{A}
  + \frac23 \bvec{A} \wedge \bvec{A} \wedge \bvec{A}).
\end{eqnarray}
The difference with the Chern-Simons 7-form studied in~\cite{B-L-N}
is that $\bvec{A}$ is not an ordinary 1-form connection,
but a superconnection including high degree of differential
forms, so that the integrand has 8-form components.
Under variations of the superconnection, we have 
\begin{eqnarray}
  \delta S = \int_M \Omega \wedge {\rm str} \{
  (d \bvec{A} + \frac12 \bvec{A} \wedge \bvec{A})
  \wedge \delta \bvec{A} \}.
\end{eqnarray}
Thus the equation
\begin{eqnarray}\label{eq:OmegaF}
  \Omega \wedge \bvec{F}= 0
\end{eqnarray}
gives a stational point of the Chern-Simons action.

We now show that the solutions of (\ref{eq:OmegaF})
describe Spin(7) holonomy manifolds under the following situation:
\begin{namelist}{(1)}
\item[(a)]
Let us choose the Euclidean space $\Bbb{R}^8$ for the manifold
$M$. Then we have the Cayley 4-form $\Omega$
given by (\ref{flat-Cayley}).
\item[(b)]
Let $\frak{sdiff}(M^8)$ be an infinite-dimensional Lie algebra
of all volume-preserving vector fields on an 8-dimensional
space $M^8$ with volume form $\omega$.
Then the vector space 
${\frak g}=\bigoplus_{p=1}^{8} \Lambda^p \frak{sdiff}(M^8)$
($\Lambda^p \frak{sdiff}(M^8)$ denotes the space of p-vectors 
$X=X_1 \wedge X_2 \wedge \cdots \wedge X_p, \, X_i 
\in \frak{sdiff}(M^8)$)
becomes a Lie superalgebra ${\frak g}={\frak g}^+\oplus {\frak g}^-$
by the Schouten-Nijenhuis bracket $[ \, , \, ]_{\rm SN}$.
We take this Lie superalgebra, i.e. $\bvec{A} \in 
{\cal A}^-(\Bbb{R}^8, {\frak g})$.
\end{namelist}
For the equation (\ref{eq:OmegaF}) we assume that the superconnection
$\bvec{A}$ is independent of the coordinates 
$x^a \, (a = 1,2,\cdots,8)$ on $\Bbb{R}^8$,
and further is of the form,
\begin{eqnarray}
  \bvec{A} = \frac12 dx^a \wedge dx^b \otimes V_{ab}
  \in {\cal A}^2(\Bbb{R}^8, {\frak g}^-)
\end{eqnarray}
where $V_{ab}=V_a \wedge V_b \in {\frak g}^-$
denote the 2-vectors on $M^8$.
Then the equation (\ref{eq:OmegaF}) is actually identical to
the 2-vector condition (\ref{eq:8MN}).
Note that the volume-preserving condition 
(\ref{eq:8vol-pre}) is
automatically satisfied by the choice of $\frak{sdiff}(M^8)$.
Thus the solution induces the Cayley 4-form on the internal space
$M^8$ with the help of the Proposition 2.

\begin{table}
\caption{Comparison of gravitational instantons}
\label{table1}
\begin{center}
\begin{tabular}{|c||c|c|}
  \hline
  && \\
  & 4-dim.  & 
  8-dim.  \\
  && \\
  \hline 
  && \\
  holonomy group & Sp(1) & Spin(7) \\
  && \\
  \hline
  && \\
  equations of motion & 
  $\begin{array}{c}
    {\bar{\eta}^a}_{\mu \nu} [V_\mu, V_\nu ] = 0 \\
    L_{V_\mu} \omega = 0
  \end{array}$ &
  $
  \begin{array}{c}
    \Omega_{\alpha \beta \gamma \delta} [V_\alpha \wedge V_\beta ,
    V_\gamma \wedge V_\delta ]_{\rm SN} = 0 \\
    L_{V_\alpha} \omega = 0
  \end{array}$ \\
  && \\
  \hline
  && \\
  Yang-Mills correspondence &
  $\begin{array}{c}
    \Sigma^a \wedge F = 0 \\
    (\Leftrightarrow F+*F=0) \\
    \frak{g}=\frak{sdiff}(M^4)
  \end{array}$ &
  $
  \begin{array}{c}
    \Omega \wedge \bvec{F} = 0 \\
    \frak{g}=\bigoplus_{k=1}^8 \wedge^k \frak{sdiff}(M^8)
  \end{array}$ \\
  && \\
  \hline
\end{tabular}
\end{center}
\end{table}

\section{Discussion}
We have provided a new approach to Spin(7) manifolds
using volume-preserving vector fields.
In table \ref{table1}, we present a summary of our results
and compare them with the analogous properties of
4-dimensional gravitational instantons (hyperk\"ahler manifolds).


The present framework may be applied to the case of 
${\rm G}_2$ holonomy manifolds or 7-dimensional gravitational
instantons.
For the ${\rm G}_2$ case, the Cayley 4-form $\Omega$ is 
replaced by a 3-form $\Phi$ and its Hodge dual 4-forms $\ast \Phi$.
Then the condition of ${\rm G}_2$ holonomy is given by 
$d \Phi =0$ and $d \ast \Phi=0$~\cite{B-S}.
Our point of view is that one must introduce a system of 
differential equations, eventually interpreted as a
certain Yang-Mills theory.
As a candidate for such equations, we propose a
pair of equations:
\begin{namelist}{}
  \item[(a) volume-preserving condition]
    \begin{eqnarray}
      L_{V_a} \omega_1 = 0, \quad L_{U_a} \omega_2 = 0
    \end{eqnarray}
  \item[(b) 2-vector condition]
    \begin{eqnarray}
      \psi_{abc} [V_a \wedge V_b , V_c]_{\rm SN} = 0, \quad
      \Omega_{abcd}[U_a \wedge U_b , U_c \wedge U_d ]_{\rm SN} = 0
    \end{eqnarray}
\end{namelist}
where the coefficients $\psi_{abc}$ and $\Omega_{abcd}$
are defined by (\ref{varphi}) and (\ref{Omega}).
Indeed, these equations reproduce the ${\rm G}_2$ holonomy
manifold given in~\cite{G-P-P} by a suitable choice
of the volume forms $\omega_1$, $\omega_2$ 
and the vector fields $V_a$, $U_a$.
The details for ${\rm G}_2$ holonomy manifolds 
will be presented elsewhere~\cite{inprep}.

Finally, we would like to mention about higher dimensional 
Ashtekar gravity.
The action of 4-dimensional Ashtekar gravity 
is the chiral form of the usual Einstein-Hilbert action.
The variables of the chiral action
consist of self-dual connection $A_+^a$ and self-dual 2-forms
$\Sigma^a \, (a=1,2,3)$.
Using these variables, we can easily read the 
Sp(1) holonomy condition, $A_+=0$.
If we bring the self-dual space into focus,
the equations of motion are reduced to $d\Sigma^a=0$.
On the other hand, in 8-dimensions, 
we know the Spin(7) holonomy condition,
which is given by $d \Omega = 0$, i.e.
$\Omega$ is the Cayley 4-form.
So we may suppose the Spin(7)-invariant self-dual
4-form $\Omega$ as an 8-dimensional Ashtekar variable,
although the precise details of the 8-dimensional Ashtekar
gravity have not been given.

\section*{Acknowledgment}
We would like to thank R. Goto, Y. Hashimoto and H. Kanno
for useful discussions.
The work of Y. Y. is supported in part by the Grant-in-Aid
for Scientific Reserch (C) No. 12640074 of the Ministry of
Education, Science, Sports and Culture of Japan.
  
\section*{Appendix \quad  Schouten-Nijenhuis bracket}

\renewcommand{\theequation}{A.\arabic{equation}}\setcounter{equation}{0}

In this Appendix, we assume $M$ is an arbitary $N$-dimensional manifold.
Let $\wedge^p TM$ denotes the space of $p$-vectors, i.e.
skew symmetric contravariant tensor fields of type $(p,0)$,
and $\wedge TM= \oplus_{p=1}^N \wedge^p TM$.
We fix a $\Bbb{Z}_2$-grading in $\wedge TM$ as follows:
the parity $|X|$ of $X=X_1 \wedge X_2 \wedge \cdots \wedge X_k 
\, (X_i \in TM)$ equals 0 or 1
according to whether $k$ is odd or even,
respectively.
A Lie superalgebra on $\wedge TM$
\begin{eqnarray}
  [\, , \, ]_{\rm SN} : \wedge^p TM \times \wedge^q TM 
  \longrightarrow \wedge^{p+q-1}TM
\end{eqnarray}
is defined by~\cite{V}
\begin{eqnarray}
  [X,Y]_{\rm SN}=\sum_{k,\,l}(-1)^{k+l}[X_k, Y_l]
  \wedge X_1 \wedge \cdots \check{X}_k
  \cdots \wedge X_p \wedge Y_1
  \wedge \cdots \check{Y}_l  \cdots \wedge
  Y_q 
\end{eqnarray}
for $X=X_1 \wedge X_2 \wedge \cdots \wedge X_p \in \wedge^p TM$, 
and $Y=Y_1 \wedge Y_2 \wedge \cdots \wedge Y_q \in \wedge^q TM$.
This operation is known as the Schouten-Nijenhuis bracket
and satisfies the axioms of a Lie superalgebra:
\begin{eqnarray}
  &[X,Y]_{\rm SN}+(-1)^{|X||Y|}[Y,X]_{\rm SN}= 0, \\
  &[X, [Y,Z]_{{\rm SN}}]_{\rm SN}=
  [[X,Y]_{{\rm SN}},Z]_{\rm SN} + (-1)^{|X||Y|}
  [Y, [X,Z]_{{\rm SN}}]_{\rm SN}. \label{eq:sJacobi}
\end{eqnarray}




\end{document}